\documentclass[12pt,a4paper,oneside]{article}
\usepackage{graphicx,amssymb,amsmath,slashed}
\usepackage{times,bbm,hyperref,color,graphicx,feynmp,multirow,appendix,latexsym,amsmath,amssymb,bm,graphics,setspace}
\usepackage[margin=2cm]{geometry}
\newcommand{\eexp}[1]{\mbox{e}^{#1}}

\newcommand{\abs}[1]{\left| #1 \right|}
\newcommand{\pd}[2]{\frac{\partial #1}{\partial #2}}

\title{\fontsize{16}{20}\selectfont\textbf{Shining Flavor and Radion Phenomenology in Warped Extra Dimension}}
\author{\fontsize{12}{16}\selectfont Yochay Eshel,$^a$ Seung J. Lee,$^{a,b}$ Gilad Perez,$^a$ and Yotam Soreq$^a$ \vspace{6pt}\\
\fontsize{11}{16}\selectfont\textit{$^a$Department of Particle Physics and Astrophysics, Weizmann Institute of Science, Rehovot 76100, Israel}\\
\fontsize{11}{16}\selectfont\textit{$^b$Department of Physics, KAIST, Daejeon 305-701, Korea}}
\date{}

\begin{document}

\maketitle

\begin{abstract}
We study radion phenomenology in the context of flavor shining in warped extra dimension models.
In this unique setup, originally proposed by Rattazzi and Zaffaroni, solutions to the gauge hierarchy problem and the new physics flavor problem are unified.
A special role is played by the vacuum energy on the branes, that naturally allows for flavon stabilization and parametrically raises the radion mass. 
We note that the radion mass squared is suppressed only by the log of the weak-Planck hierarchy, and it is in the favored range of the standard model Higgs.
We emphasize that the radion to di-photon, to $\tau\tau$ and to $WW^*$ can be promising discovery channels at the LHC, with a rate above that of the standard model Higgs.
We find that the radion is unlikely to account for the excess in $W$ plus dijet events as recently reported by the CDF collaboration. 
\end{abstract}

%%%%%%%%%%%
\section{Introduction} \label{Intro}
%%%%%%%%%%%%

The Randall-Sundrum (RS) framework consists of a slice of anti de Sitter space in five dimensions (AdS$_5$), where the warped geometry naturally generates the weak-Planck hierarchy~\cite{RS1} and offers new approaches to flavor physics \cite{Grossman:1999ra}. 
It may also address the standard model (SM) flavor puzzle via the split-fermion mechanism~\cite{ArkaniHamed:1999dc}, using flavor dependent wave function localization for the SM fermions \cite{GP_bulk_fields,Huber:2000ie}. 
In addition, a protection from large flavor and CP violation is obtained via the so called RS-GIM mechanism~\cite{GP_bulk_fields,Huber:2000ie,Huber:2003tu,Agashe:2004ay, Agashe:2004cp}, even if the fundamental flavor parameters are anarchic. 
However, a residual little CP problem, in the form of sizable contributions to $\epsilon_K$, $\epsilon'/\epsilon_K$~\cite{Bona:2007vi, Davidson:2007si, Csaki:2008zd,Blanke:2008zb,Bauer:2009cf,Gedalia:2009ws} and to the electric dipole moments~\cite{Agashe:2004ay, Agashe:2004cp, Cheung:2007bu,Delaunay:2011vv}, ruins such a nice feature. 
For further discussions on the RS flavor problems see {\it e.g.}~\cite{ Cacciapaglia:2007fw,Burdman:2002gr,Burdman:2003nt,Albrecht:2009xr,
Agashe:2006wa, Casagrande:2008hr,Agashe:2008uz,Blanke:2008yr,Csaki:2009bb,Buras:2009ka,Agashe:2005hk,Ligeti:2006pm} .
This RS little CP problem pushes the Kaluza Klein (KK) scale to be larger than $\mathcal{O}(10 \, {\rm TeV})$~\cite{Delaunay:2011vv}, which is beyond of the LHC reach and implies a severe little hierarchy problem. Gauging the SM approximate flavor symmetries (or part of them) in the bulk may solve this problem \cite{RZ}, as was recently investigated by various authors \cite{Cacciapaglia:2007fw,Csaki:2009bb, Fitzpatrick:2007sa, Csaki:2008eh, Perez:2008ee, Csaki:2008qq, Chen:2008qg, Santiago:2008vq}.

In a large class of 5D flavor models the flavor symmetry is broken by unspecified dynamics on the UV-brane, and flavor violation is shined \cite{shining,RZ,RSshining,Delaunay:2010dw} to the IR-brane by scalar flavon fields, which acquire vacuum expectation values (VEVs). 
It is interesting that a model of this type, where the full SM flavor hierarchies are generated by some mechanism on the UV-brane, can ameliorate the little hierarchy problem and improve the visibility of the model \cite{Delaunay:2011vv,Delaunay:2010dw}. 
The above possibility is also motivated by the  AdS/CFT correspondence \cite{Maldacena:1997re,Aharony:1999ti}, where the RS setup can be understood from a 4D point of view \cite{RZ,AHNPMR,Agashe:2002jx,Contino:2004vy,PerezVictoria:2001pa}, identifying 4D global currents with bulk gauge symmetries. Indeed, in the context of electroweak precision tests, a significant improved Electroweak fit is obtained when the custodial symmetry of the SM is gauged in the bulk~\cite{CUSTODIAL}. 

The gauge hierarchy problem can be addressed by the RS framework only if the size of the extra dimension is stabilized to $\mathcal{O}(40)/k$, where $k$ is the AdS$_5$ curvature.
Therefore, the radion field, which corresponds to fluctuations in the distance between the branes, should develop a VEV (and a positive mass squared) that sets the size of the extra dimension. 
The stabilization problem in the RS framework was studied by~\cite{GW1,GW2,dewolfe,TM,CGK,csaki-2004,GP,GR,Cabrer,GPT,KNQ,LS_RS1,KMP,CGR,RZ,Haba:2011tc} , when the favored solution is the Goldberger-Wise (GW) mechanism~\cite{GW1,GW2}.
It requires the presence of an additional bulk scalar field with a non-trivial background along the extra dimension, and a relatively small 5D mass.
It is interesting to unify the fields which induce shining of flavor violation and stabilization, such as the GW scalar and the flavons.
The idea of utilizing flavons as GW scalars was raised by Rattazzi and Zaffaroni~\cite{RZ} for IR localized SM model. 
Here we consider the case of~\cite{RSshining}, where the SM fermions and gauge fields propagate in the bulk, that allows one to solve the little hierarchy problem, and show that  stabilization can be naturally induced by the flavons. 

Radius stabilization is induced by the presence of flavon VEVs and brane tension terms. Furthermore, the radion mass squared is parametrically enhanced, as it is suppressed only by the log of the weak-Planck hierarchy, in agreement with the recent result of~\cite{KNQ}.  The radion mass is found to be in the favored range of the SM Higgs mass. 
Hence, in this model the radion is likely to be the lightest new particle and may be the first signal of this framework. 

The remaining of the paper is organized as follows. In section~\ref{GWStabMech} we study the GW mechanism in the presence of brane tension deviation and estimate the radion mass, while in section~\ref{shinstab} we utilize the flavon fields as GW scalars. 
Section~\ref{LHC_Discovery} contains a discussion on the collider phenomenology of the radion.  
We end with our conclusions in section~\ref{Conclusions}\,. The Appendices contain technical derivations. 

%%%%%%%%%%%%%%%%%%%%%%%%%%%%%%%%
\section{The Goldberger-Wise Mechanism and the Radion Mass}  \label{GWStabMech}
%%%%%%%%%%%%%%%%%%%%%%%%%%%%%%%%

The GW mechanism provides a classical solution to the stabilization problem of the RS framework. 
In order to ensure that the main appealing features of the RS framework are retained, we consider a small back-reaction of the GW scalar on the metric. 
The background solution of the gravity-scalar system is obtained using perturbation theory, following the method proposed in \cite{LS_RS1}. 
 For simplicity, we use the 4D effective potential method \cite{GW1,GW2}, where the extra dimension is integrated out. In Appendix~\ref{linearied_EE}, it is shown that the result for the radion mass in this method is identical to the one that is obtained in the method of \cite{CGK,KMP}.

%%%%%%%%%%%%%%%%%%%%
\subsection{The Goldberger-Wise Setup} \label{setup}

We focus on the original RS setup and the GW mechanism.
 The extra dimension, $y\in(-y_c,y_c]$, is bounded by two $3+1$ branes. One brane at $y=0$ and  is called the UV-brane, and the other one at $y=y_c$ is called the IR-brane. Since we look for a background solution that takes into account the back-reaction of the GW scalars on the metric, and preserves 4D Poincar$\acute{\text e}$ invariance, we choose the following ansatz for the metric
\begin{align} \label{metric_ansatz}
&	ds^2 =G_{MN} dx^M dx^N= \eexp{-2A(y)}\eta_{\mu\nu}dx^\mu dx^\nu -dy^2 \,.
\end{align}
Greek indices run over $(0,1,2,3)$, capital Latin for $(0,1,2,3,4)$, $\eta_{\mu\nu}={\rm diag}(1,-1,-1,-1)$ and $x_4=y$. We identify $y$ with $-y$ via orbifold, $S_1/Z_2$, assignment.
The gravity part of the action forces an AdS$_5$ background metric, with curvature $k$. It includes the Ricci scalar $\mathcal{R}$, a negative bulk cosmological constant $\Lambda_{\rm bulk}=-12M^3k^2$  and the brane tensions, which are parameterized by the dimensionless parameters  $T_{\rm IR/UV}$. The 5D Planck scale is $M\!\sim\!10^{19}{\rm\, GeV}$ and we assume that $k\lesssim 2.6M$ (see discussion on section~\ref{LHC_Discovery}\,).
 The GW mechanism contains bulk scalars, $\phi_i$, with bulk and brane potentials, which are $V^i_{\rm bulk}(\phi_i)$ and $V^i_{\rm IR/UV}(\phi_i)$ respectively. $i$ will become flavor indices later on when we consider flavon fields as GW scalars.
The considered action is 
\begin{align} \label{5Daction}
	S = \int d^5x\left(  \mathcal{L}_{\rm gravity} + \sum_{i}\mathcal{L}^{(i)}_{\rm GW}  \right) \ ,
\end{align}
where
\begin{align}
&	\mathcal{L}_{\rm gravity} = \sqrt{g}\left(12M^3k^2 - M^3\mathcal{R}\right) 
		- \sqrt{g_{\rm UV}}M^3kT_{\rm UV}\delta(y) - \sqrt{g_{\rm IR}}M^3kT_{\rm IR}\delta(y-y_c) \ , \nonumber\\
&	\mathcal{L}^{(i)}_{\rm GW}  = \sqrt{g}\left(\frac{1}{2}\nabla^M\phi_i\nabla_M\phi_i- V^i_{\rm bulk}(\phi_i) \right)  
			-\sqrt{g_{\rm IR}}V^i_{\rm IR}(\phi_i)\delta(y-y_c) - \sqrt{g_{\rm UV}}V^i_{\rm UV}(\phi_i)\delta(y) \ .  \nonumber
\end{align}
$g$ is the determinant of the metric and $g_{\rm IR/UV}$ are the determinants of the induced metric on the branes. The GW scalars potentials are
\begin{align} \label{GW_Potentials}
	\begin{array}{rl}
&	V^i_{\rm bulk}(\phi_i)= \frac{1}{2}\epsilon_i k^2 \phi^2_i \ , \\
&	V^i_{\rm IR}(\phi_i)	= \frac{1}{2}b_{{\rm IR},i} k\left(\phi_i - r\nu_{{\rm IR}, i} M^{3/2}\right)^2 \ , \\
&	V^i_{\rm UV}(\phi_i)	= \frac{1}{2}b_{{\rm UV}, i} k\left(\phi_i - r\nu_{{\rm UV}, i} M^{3/2}\right)^2 \ ,  
	\end{array}
\end{align}
where $\epsilon_i,\ b_{{\rm IR/UV},i},\ r$ and $\nu_{{\rm IR/UV},i}$ are dimensionless parameters. $\epsilon_i$ ($b_{{\rm IR/UV},i}$) parametrize the bulk (branes) masses of the GW scalars in units of the curvature and we consider $\epsilon_i$ as small parameters. 
For simplicity, we consider non-interacting quadratic potentials 
 for the GW scalars. Higher terms of the GW scalars are negligible in the perturbative expansion, for further discussion see~\cite{RZ}. 
The boundary conditions of the GW scalars are determined by the brane potentials, and can be Neumann for $b_{{\rm IR/UV},i}=0$ or modified Dirichlet for $b_{{\rm IR/UV},i}\to\infty$, where the boundaries' value of the GW scalars are set by $r$ and $\nu_{{\rm IR/UV},i}$. The result for finite values of $b_{{\rm IR/UV},i}$ interpolates between the above extremes. 
Our parameterization for  branes' potentials is chosen due to the following reason. For large $b_{{\rm IR/UV},i}$ the GW scalars VEVs on the branes are $\phi_i(0)\sim M^{3/2}r\nu_{{\rm UV},i}$ and $\phi_i(y_c)\sim M^{3/2}r\nu_{{\rm IR},i}$, which mean that the GW scalars branes' potentials contribution to the brane tensions is small in this case.
As explained in section~\ref{Background}\,, $r$ serves as the expansion parameter for the background solution. Therefore, for later use we parametrize the brane tension as follows
\begin{align}
&	T_{\rm IR} = T^{(0)}_{\rm IR}+r^2\delta{T_{\rm IR}} \ , \qquad
	T_{\rm UV} =T^{(0)}_{\rm UV} +r^2\delta{T_{\rm UV}} \ . \label{Texpansion}
\end{align}

%%%%%%%%%%%%%%%%%%%%%
\subsection{The Background Solution} \label{Background}

The background solution preserves 4D-Poincar$\acute{\text e}$ invariance, i.e. we consider solutions where the GW scalars and the metric backgrounds  are only $y$-dependent.
The metric is symmetric under the orbifold symmetry and we assume that for the GW scalars as well.
The GW scalars equation of motion and the 5D Einstein equations, which are derived from Eqs.~\eqref{metric_ansatz}-\eqref{GW_Potentials} are
\begin{subequations} \label{EEq}
\begin{align} 
	\phi''_i - 4A'\phi'_i-\epsilon k^2 \phi_i 
=&\, 	b_{\rm UV,i} k\left(\phi_i - r\nu_{{\rm UV},i} M^{3/2}\right)\delta(y) 
		+b_{{\rm IR},i} k\left(\phi_i - r\nu_{{\rm IR},i} M^{3/2}\right)\delta(y-y_c) \ , \label{EEq1}\\
	6M^3A''
=&\, 	\phi^{'2}_i + \left[ M^3kT_{\rm UV}+\frac{1}{2}b_{{\rm UV},i} k\left(\phi_i - r\nu_{{\rm UV},i} M^{3/2}\right)^2 \right]\delta(y) \nonumber\\ 
 	& \quad \ \,  
		+ \left[ M^3kT_{\rm IR}+\frac{1}{2}b_{{\rm IR},i} k\left(\phi_i - r\nu_{{\rm IR},i} M^{3/2}\right)^2   \right]\delta(y-y_c) \ , \label{EEq2}\\
	A'^2 
=&\,	 k^2 + \frac{1}{24M^3}\left( \phi'^2_i - \epsilon k^2\phi^2_i\right)  \ , \label{EEq3}
\end{align}
\end{subequations}
where repeated indices are summed over and $'$ are derivatives with respect to $y$. From Eq.~\eqref{EEq3} one can see that non zero VEVs of the GW scalars will perturb the metric away from the pure AdS$_5$ form. 
Later we will see that the background solution of the GW scalars is proportional to $r$ (and higher powers of it), therefore it naturally serves as an expansion parameter and we can ensure that the back-reaction of the metric from the GW scalars is small.

The zeroth order equations can be obtained by setting $r=0$ and $\phi_i(y)=0$. After we integrate over Eqs.~\eqref{EEq2} and~\eqref{EEq3} (where Eq.~\eqref{Texpansion} is plugged into Eq.~\eqref{EEq2}) we get
\begin{align} \label{Azero_sol}
	A(y) = k\abs{y} \ , \quad
	T^{(0)}_{\rm UV} = 12 \ , \quad
	T^{(0)}_{\rm IR} = -12 \ .
\end{align}
This is the pure AdS$_5$ solution, and the values of $T^{(0)}_{\rm IR/UV}$ are those of the unstabilized RS model. Therefore, $\delta{T}_{\rm IR/UV}$ are the brane tensions deviations. From now on, we set $A(0)=0$, which is a gauge fixing.
 In Eq.~\eqref{Azero_sol} we chose the positive solution of the square root, $A'=\abs{k}$ (the negative solution, $A'=-\abs{k}$, corresponds to interchanging the IR and UV branes). 

To first order in $r$, Eq.~\eqref{EEq1} in the bulk is
 \begin{align}
	&	\phi''_i - 4k\phi'_i-\epsilon_i k^2 \phi_i = 0 \label{GW_eom} \, ,
 \end{align} 
with the boundary conditions
\begin{subequations} \label{GW_bc}
\begin{align}
&	 2\phi'_i(0) = b_{{\rm UV},i} k\left(\phi_i(0) - r\nu_{{\rm UV},i} M^{3/2}\right) \ , \\ 
&	 2\phi'_i(y_c) = -b_{{\rm IR},i} k\left(\phi_i(y_c)- r\nu_{{\rm IR},i} M^{3/2}\right) \ .
\end{align}
\end{subequations}
The solution of Eqs.~\eqref{GW_eom}-\eqref{GW_bc} is 
\begin{align}\label{GW_background} 
	\phi_i(y) =& rM^{3/2}\left(X_{1,i}\eexp{\left(2-\sqrt{4+\epsilon_i}\right)k\abs{y}} + X_{2,i}\eexp{\left(2+\sqrt{4+\epsilon_i}\right)k\abs{y}} \right) \ , 
\end{align}
where $X_{i,1}$ and $X_{i,2}$ are given by
\begin{subequations}
\begin{align} 
	X_{i,1} 
=&  	\frac{\eexp{2ky_c\sqrt{4+\epsilon_i }} \left(2 \sqrt{4+\epsilon_i}+4+b_{\rm IR}\right)\nu_{{\rm UV},i}
	-\eexp{ky_c\left(-2+\sqrt{4+\epsilon_i}\right)}b_{{\rm IR},i}\nu_{{\rm IR},i}}
		{\eexp{2ky_c\sqrt{4+\epsilon_i}} \left(2\sqrt{4+\epsilon_i}+4+b_{\rm IR}\right)+2 \sqrt{4+\epsilon_i}-b_{\rm IR}-4} \ ,  \\
	X_{i,2} 
=&	\nu_{{\rm UV},i} - X_{1,i} \ . 
\end{align}
\end{subequations}
and consider the the limit of $b_{{\rm UV},i}\to\infty$. As long as $b_{{\rm UV},i}\gg\epsilon_i$, taking finite $b_{{\rm UV},i}$ does not change our results. As we claimed before, $\phi_i\propto r$ therefore $A(y)$ and $T_{\rm UV/IR}$ receive no contributions at first order in $r$.

To order $r^2$ the GW scalars' equations are left unchanged, but the one involving $A(y)$ is
\begin{align} 
&	A'(y) = k + \frac{1}{48M^3k}\left(\phi'^2_i(y) - \epsilon k^2\phi^2_i(y)\right)   \label{Atag} \ ,
\end{align}
and after integration we get 
\begin{align}
&	A(y) = k\abs{y} + r^2kG\left(\abs{y}\right) \ , \label{Asol} 
\end{align}
where
\begin{align}
	G(y) = \int_0^{y}\frac{d\alpha}{12} \eexp{2k\alpha \left(2-\sqrt{4+\epsilon_i}\right)} \Big\{&
	\eexp{4k\alpha\sqrt{4+\epsilon_i}}X_{2,i}^2 \left(2+\sqrt{4+\epsilon_i}\right) \nonumber\\
&	-\eexp{2k\alpha\sqrt{4+\epsilon_i}}X_{1,i}X_{2,i} \epsilon_i + X_{1,i}^2 \left(2-\sqrt{4+\epsilon_i}\right)\Big\} \ . 
\end{align} 
$G(y)$ is the leading order correction to the AdS$_5$ metric from the GW scalars back-reaction.  

The expansion in power series of $r$ naively requires $r\ll1$, but in the present case one can relax this condition. In fact, for a sizable range of parameter space, discussed below, $r$ can be of order unity. 
From Eq.~\eqref{EEq3} we find that 
\begin{align}
	A' = k\sqrt{1+\frac{\phi'^2_i-\epsilon k^2\phi^2_i}{24M^3k^2}} \ .
\end{align} 
 In order to have an AdS$_5$-like solution we expand the square root in the above equation to lowest non-trivial order. The condition for the validity of this approximation is 
\begin{align} \label{r_cond_2}
	\abs{\phi'^2_i - \epsilon k^2\phi^2_i}\ll 24k^2M^3 \ ,
\end{align}
which, as anticipated, allows for $r\sim\mathcal{O}(1)$ for a wide range of the parameters. 

%%%%%%%%%%%%%%%%%%%%%%%%%%%%%
\subsection{The 4D Effective Action and the Radion Mass} \label{Effective_potential}

The 4D effective action is derived by integrating out the fifth dimension, similarly to \cite{GW1}, for detailed derivation see Appendix~\ref{Deriv_R4DEA}\,. 
Let us denote the radion field as $\varphi_R(x^\mu)=fa(x^\mu)$, where 
\begin{align} \label{af_def}
	a(x^\mu)\equiv\eexp{-ky_c(x^\mu)} \ , \qquad \qquad \qquad
	f\equiv\sqrt{12\frac{M^3}{k}} \ .
\end{align}
For successful stabilization the radion VEV should produce the weak-Planck hierarchy, therefore $\langle a \rangle\equiv a_\star\sim10^{-16}$.
The effective action to order $r^2$  and leading order in $\epsilon$ and $a$ is
\begin{align} \label{L_4D_GW}
	\mathcal{L}^{\rm 4D}_{\rm eff} 
=&	\frac{1}{2} f^2\left(\partial_\mu a\right)^2 - V_{\rm eff}(a) + const \ , 
\end{align}
where
\begin{align}
	V_{\rm eff}(a)  \label{VGW}
=	r^2kM^3a^4\Bigg\{& \delta{T}_{IR}
	+\sum_i\Bigg[ \frac{4b_{{\rm IR},i}}{8+b_{{\rm IR},i}} (\nu_{{\rm UV},i}a^{\epsilon_i/4}-\nu_{{\rm IR},i} )^2
		+\frac{\epsilon_i}{4(8+b_{{\rm IR},i})^2}\Big[\left(b_{{\rm IR},i}\nu_{{\rm IR},i}\right)^2 \nonumber\\
&		+2a^{\epsilon_i /2}\nu_{{\rm UV},i}^2\left(b_{{\rm IR},i}^2-32\right) 
		- 4a^{\epsilon_i /4}\nu_{{\rm UV},i}\nu_{{\rm IR},i} b_{{\rm IR},i}(4+b_{{\rm IR},i})  \Big] \Bigg]   \Bigg\}  \, .
\end{align}
The constant in Eq.~\eqref{L_4D_GW}, sets the value of the potential to zero at its minimum, by tuning the value of $\delta{T_{\rm UV}}$. The corresponding fine-tuning in this procedure is nothing but the celebrated cosmological constant problem, whose solution is beyond the scope of this work. Eq.~\eqref{VGW} gives the leading order form of the effective potential, we note that $G(y)$ contributes only to order $r^4$ of the effective potential.
 
The radion mass is derived by using the effective potential. For simplicity we consider the case of one GW scalar, and denote $\epsilon_i = \epsilon$, $b_{{\rm IR},i}=b_{\rm IR}$, $\nu_{{\rm UV},i}=1$ and $\nu_{{\rm IR},i}= \nu$. 
To leading  order in $\epsilon$ and $1/\log(a_\star)$, the extremum condition, $\partial V/\partial a\big|_{a=a_\star}\!\!\!\!\!\!\!\!\!\!=0$, is  
\begin{align} \label{a_term}
	a^{\epsilon/4}_{\star\pm} 
= 	\nu \pm \frac{\sqrt{-\delta T_{\rm IR}b_{\rm IR}(8+b_{\rm IR})}}{2b_{\rm IR}} + \mathcal{O}\left(\epsilon\right)  \ ,
\end{align}
which are the two extrema to of the effective potential.
The radion mass is 
\begin{align} \
	m^2_{\rm rad,\pm} 
=&	 \frac{1}{f^2}\pd{^2V(a)}{a^2}\Bigg|_{a=a_{\star\pm}}\!\!\!\!\!\!\!\!\!\! \nonumber\\
=& 	\pm\frac{1}{3}k^2a^{2}_{\star\pm} \epsilon r^2 \frac{\sqrt{-\delta T_{\rm IR}b_{\rm IR}(8+b_{\rm IR})}}{8+b_{\rm IR}} 
	\left( \nu \pm \frac{\sqrt{-\delta T_{\rm IR}b_{\rm IR}(8+b_{\rm IR})}}{2b_{\rm IR}}  \right)
	+ \mathcal{O}\left(\epsilon^2\right) \ , \label{mrad}
\end{align}
where the $\pm$ refer to the different solutions of $a_{\star,\pm}$ respectively. From Eq.~\eqref{mrad} we can see that one extremum corresponds to a minimum of the potential and the other to a maximum. 
For $\epsilon>0$ and $\delta T_{\rm IR}<0$ the minimum is at $a_{\star+}$ for $b_{\rm IR}>0$ and at $a_{\star-}$ for $b_{\rm IR}<-8$. For $\delta T_{\rm IR}>0$ and $-8<b_{\rm IR}<0$ the minimum is located at $a_{\star+}$. For $\epsilon<0$ the locations of the maximum and the minimum are reversed. The desired hierarchy can be achieved without fine-tuning of $\nu$, $b_{\rm IR}$ and $\delta{T}_{\rm IR}$.

Unlike previous works but in agreement with \cite{KNQ}, we find that the parametric dependence of the radion mass  is $m^2_{\rm rad}\sim k^2a^2_\star \epsilon$, and not $m^2_{\rm rad}\sim k^2a^2_\star\epsilon^{3/2}$ as in \cite{GW2}\,\footnote{In \cite{GW2} the radion mass squared that is given in the main text is $m^2_{\rm rad}\sim k^2a^2_{\star}\epsilon^2$. However as the authors mentioned in footnote~2, a careful calculation yields $m^2_{\rm rad}\sim k^2a^2_{\star}\epsilon^{3/2}$.}
 or $m^2_{\rm rad}\sim k^2a^2_\star\epsilon^2$ as in \cite{CGK}, where the IR-brane tension is tuned. 
 This parametric enhancement is due to the presence of IR-brane tension terms, and this is a general consequence of the GW mechanism.
The original result of \cite{GW2}, where $m^2_{\rm rad}\sim k^2a^2_\star\epsilon^{3/2}$, can be easily reproduced in the limit that the IR-brane tension is fine tuned $\left(\abs{\delta T_{\rm IR}}\!\ll\!\epsilon\right)$. From Eq.~\eqref{a_term} one can see that $\epsilon\sim1/\log(a_\star)$; thus, the radion mass square is  inversely suppressed by log of the weak-Planck hierarchy $\left(\log(a_\star)\right)$. For typical values of parameters we get that 
\begin{align}
	m_{\rm rad} = \mathcal{O}(100\, {\rm GeV})\, .
\end{align}
Furthermore, our result is not affected by the Casimir energy which is induced by bulk fields, see Appendix~\ref{Casimir_potential}. 

We can get some intuition for the radion mass estimation by invoking the AdS/CFT correspondence.
The conformal field theory (CFT) is defined below some UV cutoff, $\sim k$, and is broken spontaneously at some low energy scale, $f a_{\star}\sim ka_{\star}$~\cite{RZ,AHNPMR}.
The radion corresponds to a 4D dilaton, the Goldstone boson of the broken conformal symmetry,
and each 5D bulk field has a dual operator in the 4D CFT. The 4D dual of the GW scalar is a scalar operator with scale dimension of $4+\epsilon/4$ for $\abs{\epsilon}\ll1$. This operator explicitly breaks the conformal symmetry. Therefore conformal invariance is only an approximation and the dilaton becomes a pseudo-Goldstone boson, with a small mass. The same happens in the 5D theory, where the radion acquires a mass due to the GW mechanism. 
The mass  square of the dilaton is proportional to the IR breaking scale times the parameter that explicitly breaks the symmetry. This parameter is proportional to the GW scalar mass and is given by $\epsilon/4$. Therefore, naively the dilaton mass is  $m^2_{\rm dil} \sim  k^2a^2_{\star}\epsilon$, which agrees with the result in Eq.~\eqref{mrad} up to order one coefficients. The same scaling of the dilaton mass can be also obtained by trace anomaly matching \cite{Schwimmer:2010za}.

%%%%%%%%%%%%%%%%%%%%%%
\subsection{Stabilization and Fine-Tunings } \label{Stab_FT}

As discussed in the introduction, the gauge hierarchy problem can be addressed by RS only in the presence of stabilization mechanism, which reduces the number of required fine-tunings from two in the unstabilized RS to one. The remaining one corresponds to the 4D cosmological constant problem~\cite{GW1, LS_RS1,CGR}, which is left unsolved.
In the absence of stabilization, the radion 4D-effective potential is
\begin{align}
	V_{\rm eff}(a) = kM^3T_{\rm UV} + \frac{\Lambda_{\rm bulk}}{k} + a^4\left(kM^3T_{\rm IR} - \frac{\Lambda_{\rm bulk}}{k}\right) \ .
\end{align}
The two fine-tunings correspond to
 (i) having a flat potential for the radion, otherwise the radion's VEV will be driven to zero or to infinity depending on the sign of the  $a^4$ term; 
  and to (ii) the vanishing of the 4D effective cosmological constant, $V_{\rm eff}(a_\star)=0$.
The resulting conditions are
\begin{align}
	T_{\rm IR} = T_{\rm IR}^{(0)} =  \frac{\Lambda_{\rm bulk}}{k^2M^3}=-12 	\ , \  \qquad
	T_{\rm UV}= T_{\rm UV}^{(0)}  = - \frac{\Lambda_{\rm bulk}}{k^2M^3}=12 	\ .
\end{align} 
However, a stabilization mechanism results in generating new terms in the effective potential. These lead to stable configuration with non trivial VEV. Consequently, the need for the first fine-tuning is eliminated.
The second fine-tuning, that corresponds to the vanishing of the 4D-effective cosmological constant, is still unavoidable. 

The same is also manifested in the 5D picture, where the warp factor has to satisfy the jump conditions. 
They are derived by integrating of Eq.~\eqref{EEq2} over a small interval around the branes, 
\begin{subequations}
\begin{align} 
&	G'(y_c)  = -\frac{\delta{T}_{\rm IR}}{12}-\frac{2\phi'^2_i(y_c)}{12r^2M^3k^2b_{{\rm IR},i}}     \ , \label{JumpGIR} \\
&	G'(0) = \frac{\delta{T}_{\rm UV}}{12}  \ .    \label{JumpGUV}
\end{align}
\end{subequations}
It is straightforward to show that  the value of $y_c$, which is obtained by the IR-jump condition in Eq.~\eqref{JumpGIR}, is equivalent to the value of $y_c$ that minimizes the radion 4D-effective potential and that the UV-jump condition from Eq.~\eqref{JumpGUV} is equivalent to the demand of the potential vanishing at its minimum. 
In the absence of the GW mechanism $\left(\phi(y)=G(y)=0\right)$, the two jump conditions do not depend on $y_c$ therefore $\delta{T}_{\rm IR/UV}\!=\!0$ is required. This corresponds to both of the above fine-tunings in the RS-framework.
 On the other hand, when adding the GW mechanism, the jump conditions depend on $y_c$. Thus, manipulation of the value of $y_c$ to fulfill the IR-jump condition is possible, which allows also for brane tension deviation. Once $y_c$ is set, there is no such freedom in the UV-jump condition which is satisfied by fine-tuning of the UV-brane tension, again resulting with a single fine-tuning at the end \cite{LS_RS1}.

%%%%%%%%%%%%%%%%%%%%%%%%%%%%%%%%%
\section{Flavon Stabilization, Flavons as Goldberger-Wise Scalars} \label{shinstab}
%%%%%%%%%%%%%%%%%%%%%%%%%%%%%%%%%

In this section we focus on the model of~\cite{RSshining} and concentrate on the quark sector only. 
In that model, the SM fermions and gauge fields propagate in the 5D bulk, while the SM Higgs is IR localized. 
An $SU(3)_{Q}\times SU(3)_{d}$ subgroup of the SM flavor symmetry is gauged in the bulk and it is broken by bi-fundamental flavon field, $y_d$.
Although we focus on a specific model, our result also applies to other RS shining models.
 For example, in the model of~\cite{Delaunay:2010dw}, one can consider, for instance, the case of a small bottom Yukawa, where the main contribution to stabilization comes from the top flavon sector (then the bulk field is $y_u$ corresponding to up-type Yukawa and not $y_d$). In addition, our analysis can be extended to other models where both up and down Yukawa propagate in the bulk, such as \cite{Delaunay:2010dw} with sizable bottom Yukawa, where the main contributions would be from the third generation sector. In these cases one expects an even richer potential and stabilization options to arise.

The quantum numbers of the 5D counterparts of the SM doublets and down-type singlet quarks under the flavor group are $\Psi_Q\sim(3,1)$ and $\Psi_d\sim(1,3)$, while the flavon quantum numbers are $y_d\sim(3,\bar3)$. 
The 5D fermions bulk masses, $c_X$ (in units of curvature),  and the dimensionless Yukawa, $Y_d$, are to lowest order in the flavons VEV, $\langle y_d\rangle$ 
\begin{subequations} \label{5DMFVcoef}
\begin{align}
	&c_Q=  \alpha_Q \mathbbm{1}+ \tilde{\beta}_Q  \langle y_d  \rangle  \langle y_d^{\dagger}  \rangle 
		= \alpha_Q \mathbbm{1}+ \beta_Q Y_d Y_d^{\dagger} \,, \\
	&c_d= \alpha_d \mathbbm{1} + \tilde{\beta}_d \langle y_d^{\dagger}\rangle\langle y_d\rangle = \alpha_d \mathbbm{1} + \beta_d Y_d^{\dagger} Y_d \,,\\
	& k \lambda_{d} \langle y_{d}\rangle= Y_{d} \,.
\end{align}
\end{subequations}
We assume that the flavor symmetry is not broken explicitly in the bulk. Therefore, to lowest order, the bulk action for the flavon is
\begin{align}
	S_{\rm bulk}=\int d^5x \sqrt{g} \left[ {\rm Tr}[G^{MN} (D_M y_d)^{\dagger}( D_N y_d)] - \epsilon k^2 {\rm Tr}[ y_d^{\dagger}y_d] \right] \,,
\end{align}
where $\epsilon$ is dimensionless and parametrizes the bulk mass of the flavon fields and we assume $\epsilon \ll 1$ as in section \ref{GWStabMech}. It also ensures that the RS flavor solution is not ruined, as for small $\epsilon$ the flavon profile along the extra dimension will be mostly flat, see Eq.~\eqref {GW_background}. 
It is straightforward to show that for small $\epsilon$ the flavor structure of bulk RS models retains its qualities even with non-constant bulk fermion masses resulting from the flavon fields \cite{RSshining}.  

As discussed, the flavor symmetry is broken only on the UV-brane, such that  the UV boundary condition for the equation of motion is   
\begin{align} \label{flavonUVvalue}
	{D}^{\rm(diag)} \equiv
 	y_d\big|_{y=0}=\frac{M^{3/2} r}{\sqrt{2}}\left(
		\begin{array}{ccc}
			\nu_{{\rm UV},1} &0 &0\\
			0 &\nu_{{\rm UV},2} &0\\
			0 &0 &\nu_{{\rm UV},3}
		\end{array} 
	\right)  \,.
\end{align}
We have parametrized the fields' values through the dimensionless parameters $\nu_{{\rm UV}, i}$, which one can normalize by defining $\nu_{\rm UV,3}=1$, and define $\tilde{\nu}_{\rm UV}=\sqrt{\left(\nu_{{\rm UV},1}\right)^2+\left(\nu_{{\rm UV},2}\right)^2+\left(\nu_{{\rm UV}, 3}\right)^2}$. We assume that all of the entries are real. For the anarchic case, all the entries are of the same order of magnitude and therefore   $\tilde{\nu}_{\rm UV}\simeq \sqrt{3}$, while for hierarchical cases the main contribution will come from the third generation entry, and thus $\tilde{\nu}_{\rm UV}\sim1$.
Given the naive dimensional analysis estimates for the values of the flavons from \cite{RSshining}, and taking into account the factor two tuning of the $\langle y_d\rangle$ found there, we find
\begin{align} 
	r\sim \frac{\sqrt{2}}{\pi}\left(\frac{\mu}{M}\right)^{3/2}, 
\end{align}
where $\mu\lesssim M$ is the scale at which the theory becomes strongly coupled. Therefore, $r\lesssim 0.4$ and the perturbative expansion, developed in section~\ref{Background}, is expected to be valid.  
The UV boundary conditions can be formally written using the $b_{\rm UV}\to \infty$ limit of the UV brane potential
\begin{align}
	V_{\rm UV}(y_d)= b_{\rm UV} k \, {\rm Tr}\left[\left|y_d-{D^{\rm (diag)}}\right|^2\right] \ . \label{UVpot}
\end{align}
We assume that on the IR-brane the flavor symmetry is not broken explicitly. Therefore, the lowest order term which one could write is a brane mass term
\begin{align}
	V_{\rm IR}(y_d)= b_{\rm IR} k \, {\rm Tr}\left[y_d y_d^{\dagger}\right] \,.
\end{align}
This is the only relevant term one could write for the brane potential, invariant under the flavor symmetry. 

Once the potentials are specified, the background solution of the gravity-scalars system can be calculated perturbatively in $r$ as in section~\ref{Background}\,. The complex fields $y_{d}^{ij}$ are separated  to real and imaginary components $y_d^{ij}=\frac{1}{\sqrt{2}}\left( y_{Rd}^{ij}+i y_{Id}^{ij}\right)$. In the notations of section \ref{Background}, the flavons' IR-brane potential corresponds to $\nu_{{\rm IR},i}=0$ for all fields, while in the UV-brane potential $\nu_{{\rm UV},i} \neq 0$ for the real diagonal entries at the limit $b_{{\rm UV},i} \to \infty$ ($\nu_{{\rm UV},i}=0$ for all the other fields).  
The resulting background solution vanishes for fields with zero UV boundary condition, which are the off-diagonal fields and imaginary diagonal fields, $y_{Rd}^{ij}(y)=0$ for $i\neq j$ and $y_{Id}^{ij}(y)=0$ for $\forall \left(i,\,j\right)$.
For the real diagonal fields, the solution is
\begin{align} \label{FlavonProfile}
	&y^{ii}_{Rd}(y) 
= 	M^{3/2}   r \nu_{{\rm UV},i} \left(X_1\eexp{(2-\sqrt{4+\epsilon})k\abs{y}} + X_2 \eexp{(2+\sqrt{4+\epsilon})k\abs{y}} \right) \,, \\
	&X_1
=	\frac{e^{2 k y_c \sqrt{4+\epsilon }}  \left(4+2 \sqrt{4+\epsilon }+b_{\rm IR} \right)}{e^{2 k y_c \sqrt{4+\epsilon }} 
	\left(4+2 \sqrt{4+\epsilon }+b_{\rm IR} \right)-4+2 \sqrt{4+\epsilon }-b_{\rm IR}}  \, ,\
	 X_2=1-X_1\,, \nonumber
\end{align}  
which is the result given by Eq.~\eqref{GW_background} in the limit of $\nu_{{\rm IR},i} \to 0\,$. 
The radion effective potential is calculated by integrating out the extra dimension. Since the three fields have the same profile but it is rescaled by different values on the UV-brane, and because of the fact that the effective potential is the result of integrating only terms quadratic in the fields, the effective potential is simply Eq.~\eqref{VGW}, by setting $\nu_{{\rm IR},i}=0$. The result is
\begin{align} \label{ShiningPotential}
	V_{\rm eff}(a)
=	 &\,M^3 k {r}^2 a^4 \left[  \delta T_{\rm IR}   +\tilde{\nu}_{\rm UV}^2 \frac{4 b_{\rm IR}  }{8+b_{\rm IR} }  a^{\epsilon/2} \right]  
		+ \mathcal{O}(\epsilon,a^8)\,,
\end{align}
where corrections from back-reaction of order $r^4$ have been neglected. 

In contrast to the general case of GW mechanism, this potential has only one extremum. It is important to note that flavon stabilization may occur only  with an IR-brane tension deviation term (notice that these results can change when sub-leading irrelevant operators on the IR brane are considered).
 The potential extremum, which is at $a=a_\star$, and the radion mass to leading order in $\epsilon$ and $a_{\star}$ are given by
\begin{align} \label{shinmrad}
	a_{\star}^{\epsilon/2} 
=	-\frac{\delta T_{\rm IR}}{ \tilde{\nu}_{\rm UV}^2}\frac{ (8+b_{\rm IR} )}{4 b_{\rm IR}}+ \mathcal{O}(\epsilon) \ , \quad
%	m^2_{\rm rad}  
%=	\frac{2}{3}k^2 a_{\star}^{2+\frac{\epsilon}{2}}\epsilon r^2  \frac{\tilde{\nu}_{\rm UV}^2b_{\rm IR}}{(8+b_{\rm IR} )}    
%	+ \mathcal{O}(\epsilon^2) \,,
	m^2_{\rm rad}  
=	-\frac{1}{6}k^2 a_{\star}^{2}\epsilon r^2  \frac{}{}    
	\delta T_{\rm IR}
	+ \mathcal{O}(\epsilon^2) \,,
\end{align}
where we assume $b_{\rm IR}\gg\epsilon$. 
Because $a_{\star}^{\epsilon/2}\sim \mathcal{O}(1)$ for the desired hierarchy and small bulk mass, it may be inferred from Eq.~\eqref{shinmrad} that as in the original GW mechanism, the desired hierarchy can be reached without fine-tuning of $b_{\rm IR}$ and $\delta T_{\rm IR}$. The radion mass scales as in the single GW scalar case considered above, with typical values of order $\mathcal{O}(100\, {\rm GeV})$.

Note that for $b_{\rm IR} \to 0 $, which is equivalent to imposing Neumann boundary conditions on the IR-brane, the mass vanishes regardless of the sign of $\epsilon$.\footnote{\,  This result is quite general, and in fact one can show, using the effective potential method, that for GW scalars with Dirichlet and Neumann boundary conditions on the UV and IR branes respectively, there cannot be stabilization with the desired hierarchy for all values of $\epsilon$ (with no assumption on its size) even with the inclusion of the IR-brane tension deviation.}
 Careful re-derivation of the radion mass, to second order in $\epsilon$, shows that in order to have a positive mass squared, one needs $b_{\rm IR} \geq \epsilon/2$ or  $b_{\rm IR} \leq -8+ \epsilon/2$ for  $\epsilon> 0$ and $-8 - |\epsilon|/2 <b_{\rm IR}< -|\epsilon|/2$ for $\epsilon<0$. We can claim, therefore, that a necessary condition for stabilization in the order we work in ${r}$ is that there is a non-vanishing IR-brane potential for the flavons.  

There are further constraints on the parameter space from flavor physics.  
The profile of the flavon fields depends on the values of $b_{\rm IR}$ and $\epsilon$. For small positive $\epsilon$ the fields' value decreases slightly in the bulk when moving away from the UV-brane, while it increases for negative $\epsilon$. Its value on the IR-brane, which determines the 4D fermion masses, is determined largely by the value of $b_{\rm IR}$. For $b_{\rm IR} \to \pm \infty$, the fields' value on the IR brane vanishes, which is problematic for models in which the Higgs is an IR brane localized field.
Conversely, for $b_{\rm IR}\simeq -8$ the fields' value on the IR-brane diverges. This signals a breakdown of the perturbative expansion, and we expect that higher order terms in the brane potentials will be important for this parameter choice. 

One should note that gauge symmetries in the bulk and their breaking on the boundaries have a dual description using the AdS/CFT correspondence. Gauge symmetries in the bulk of unsliced AdS correspond to global symmetries of the dual CFT. Adding the UV and IR branes means that the bulk gauge fields can be decomposed to zero and higher  KK modes. The higher KK modes are interpreted as composite states of the strongly coupled CFT. While the existence of massless modes of the gauge fields means that the symmetry is also weakly gauged in the CFT. If the symmetry is broken in the gravity side by boundary conditions the zero mode will pick up a mass. If the mass is of the order of the CFT's UV cutoff, the zero mode decouples and the symmetry is in fact a global symmetry on the 4D side; on the other hand, if the mass is of the order of the IR cutoff, the interpretation is that the CFT breaking also dynamically breaks the symmetry. Therefore, imposing the flavor symmetry and breaking it on the UV-brane, as we do, implies that the flavor symmetry of the SM is imposed as a global symmetry of the CFT.  

The flavon fields are then interpreted as dual to scalar operators on the conformal side which have non trivial quantum numbers under the flavor symmetry. If the flavons have small bulk masses, adding these operators explicitly breaks the conformal symmetry only by marginally relevant or irrelevant operators. The pre-factors of the two exponents in the solution of the scalar equation of motion are related to the source of the CFT operator and its VEV respectively \cite{AHNPMR}. If an IR-brane potential for the flavon field is not added, one finds that the solution is nearly constant in the bulk and that the coefficient of the exponent proportional to the VEV is negligible, implying that inner CFT dynamics do not break the flavor symmetry. On the other hand, turning on only the brane mass terms, as we did, also turns on the coefficient of the exponent proportional to the VEV of the scalar, signaling that now the CFT breaks the flavor symmetry dynamically. However, because in this case the flavon VEVs do not twist in flavor space while propagating in the bulk \cite{Grossman:2004rm}, one interprets that the CFT dynamical breaking is aligned with the explicit breaking of the flavor symmetry, hence no additional flavor violation is induced.

%%%%%%%%%%%%%%%%%%%%
\section{Collider Phenomenology } \label{LHC_Discovery}
%%%%%%%%%%%%%%%%%%%%

We focus on the case of light radion, which in the CFT description can be viewed as a pseudo Goldstone boson of the broken conformal symmetry. In this case, as discussed above, the typical value of the radion mass is $\mathcal{O}(100\, {\rm GeV})$,
which interestingly coincides with the preferred range of  the SM Higgs mass~\cite{EW}.  
We compare the radion signal to that of the SM Higgs in three interesting channels, $gg\to r/h_{\rm SM}\to\gamma\gamma$, $gg\to r/h_{\rm SM}\to\tau^+\tau^-$ and $gg\to r/h_{\rm SM}\to WW^*$. We emphasize that in the first channel the radion signal can be naturally enhanced by a factor of $\mathcal{O}(10)$. In the two other channels the radion signal can be larger by a factor of $\mathcal{O}(5)$ compared to the SM Higgs case.  
 In addition, we show that the radion is unlikely to account for the excess in the $W$ plus dijet events as recently reported by the CDF collaboration \cite{CDFWjj,CDFWjjweb}. It is important to note that in an equivalent study of the D0 collaboration there is no excess \cite{Abazov:2011af}.

The radion collider phenomenology has been discussed in details by~\cite{GW2,CGK,CGR,Rattazzi_CL,GGS,Bae,Dominici,Gunion,GRW,Rizzo,Kribs_TASI,CHS,Toharia,Davoudiasl:2009cd,Brooijmans:2010tn,Cheung:2000rw}.
Here we focus some of the important features relevant to our analysis, especially in view of the LHC improved sensitivity to the above class of observables. 
The radion couplings to the matter fields can be obtained from general principles, similar to the known bonafide dilaton.
 The couplings are proportional to the mass of the fields (or more precisely to the effective 4D trace of the energy momentum tensor when integrated over the extra dimension~\cite{CHS}),
\begin{equation}
g^i_{\rm rad}\propto m_i/\Lambda_{\rm rad}\,,
\end{equation}
where $i$ stands for a generic matter field,
and $\Lambda_{\rm rad}$ is defined as
\begin{align}
	\Lambda_{\rm rad}=\sqrt{12\frac{M^3}{k}} a_\star \ .
\end{align}
$\Lambda_{\rm rad}$ is the most important parameter for the radion phenomenology,
and can be interpreted as a symmetry breaking scale in the 4D CFT language~\cite{CHS}.
The value of $\Lambda_{\rm rad}$ is roughly around $\mathcal{O}(\,{\rm TeV})$,
but it can be viewed as a free parameter of the theory as long as it satisfies the perturbative bound~\cite{Chacko:1999hg,Agashe:2007zd}. The perturbative limit implies that $k/M_{\rm Pl}\lesssim 3$, where $M_{\rm Pl}$ is the 4D Planck scale. From the relation, $M^2_{\rm Pl}\approx2M^3/k$ (the notation of~\cite{CGK} is used), we obtain that $k/M \lesssim 2.6$. 
Therefore, for $ka_\star\sim\mathcal{O}(1\,{\rm TeV})$, there is no problem to consider $\Lambda_{\rm rad}\sim\mathcal{O}(1\,{\rm TeV})$.
  
In the $\mathcal{O}(100\,{\rm GeV})$ mass range, just as in the SM Higgs case, the di-photon discovery channel is one of the most important channels for the radion search at the LHC.
In fact, it is found that the radion discovery potential in the $gg\to r\to\gamma\gamma$ process can be enhanced compared to $gg\to h_{\rm SM}\to\gamma\gamma$
for the reasonable theory parameter space~\cite{CHS}.
The reason for the enhancement is as follows. The dominant contribution to the SM Higgs production via the gluon fusion involves triangle loop diagram with heavy fermions, and therefore, is proportional to the QCD beta function coefficient of the top quark ($\rm{b}_t$)~\cite{Djouadi:2005gi}. However, the dominant contribution to the gluon fusion production for the radion comes from the trace anomaly, i.e., it is (in the CFT language) proportional to the beta function coefficient of massive composite fields ($\rm{b}_{\rm{CFT}}$)~\cite{CHS}. Because $\rm{b_{CFT}}\gg\rm{b}_t$, it follows that the radion coupling to a pair of gluons can be larger than that of the SM-Higgs. On the other hand, the branching ratio of the radion to di-photon is similar to that of the SM Higgs in most of the parameter space~\cite{CHS}.
 
\begin{figure}[htbp]
\centerline{
	\includegraphics[width=.5\hsize]{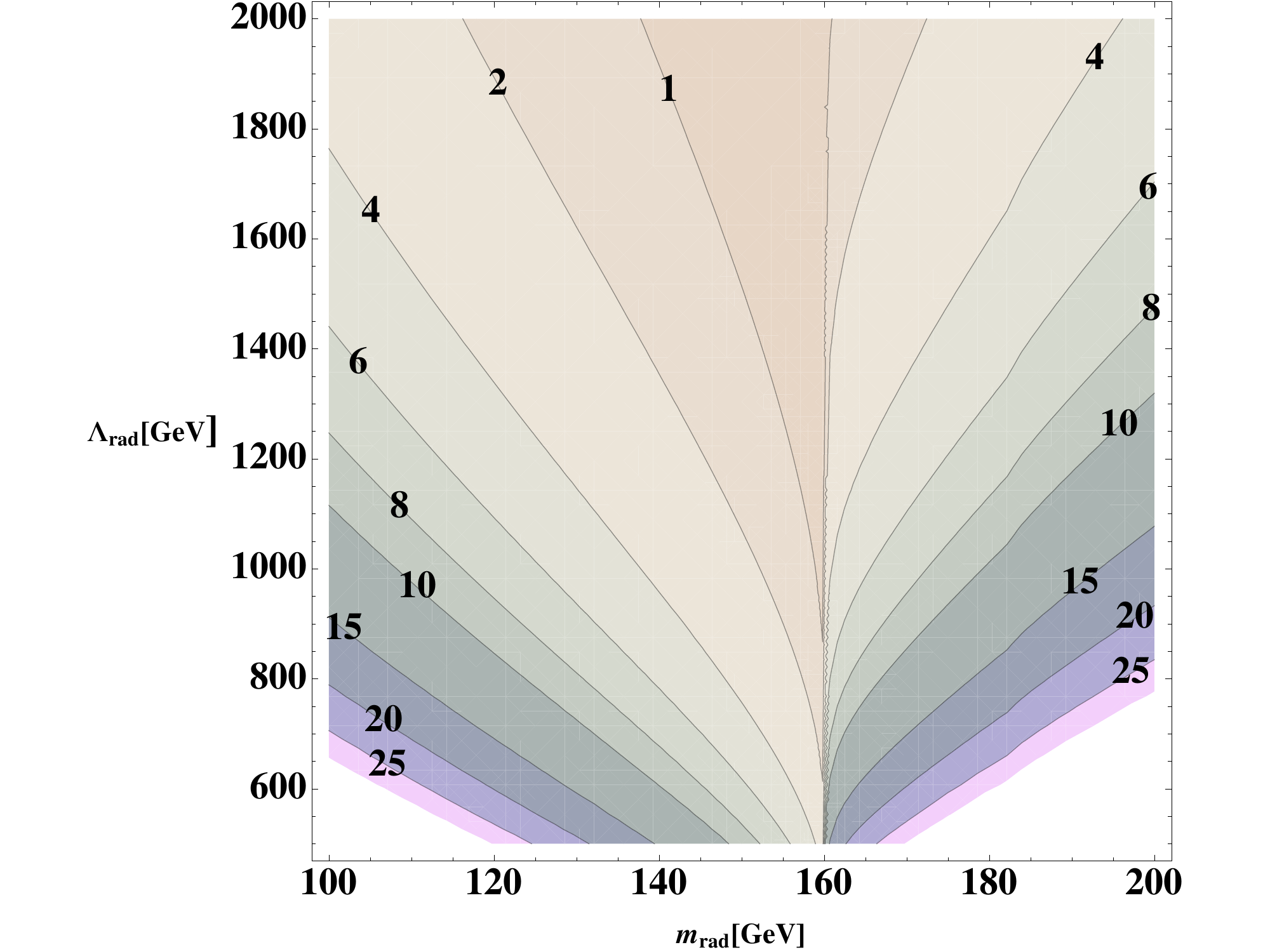}
	\includegraphics[width=.5\hsize]{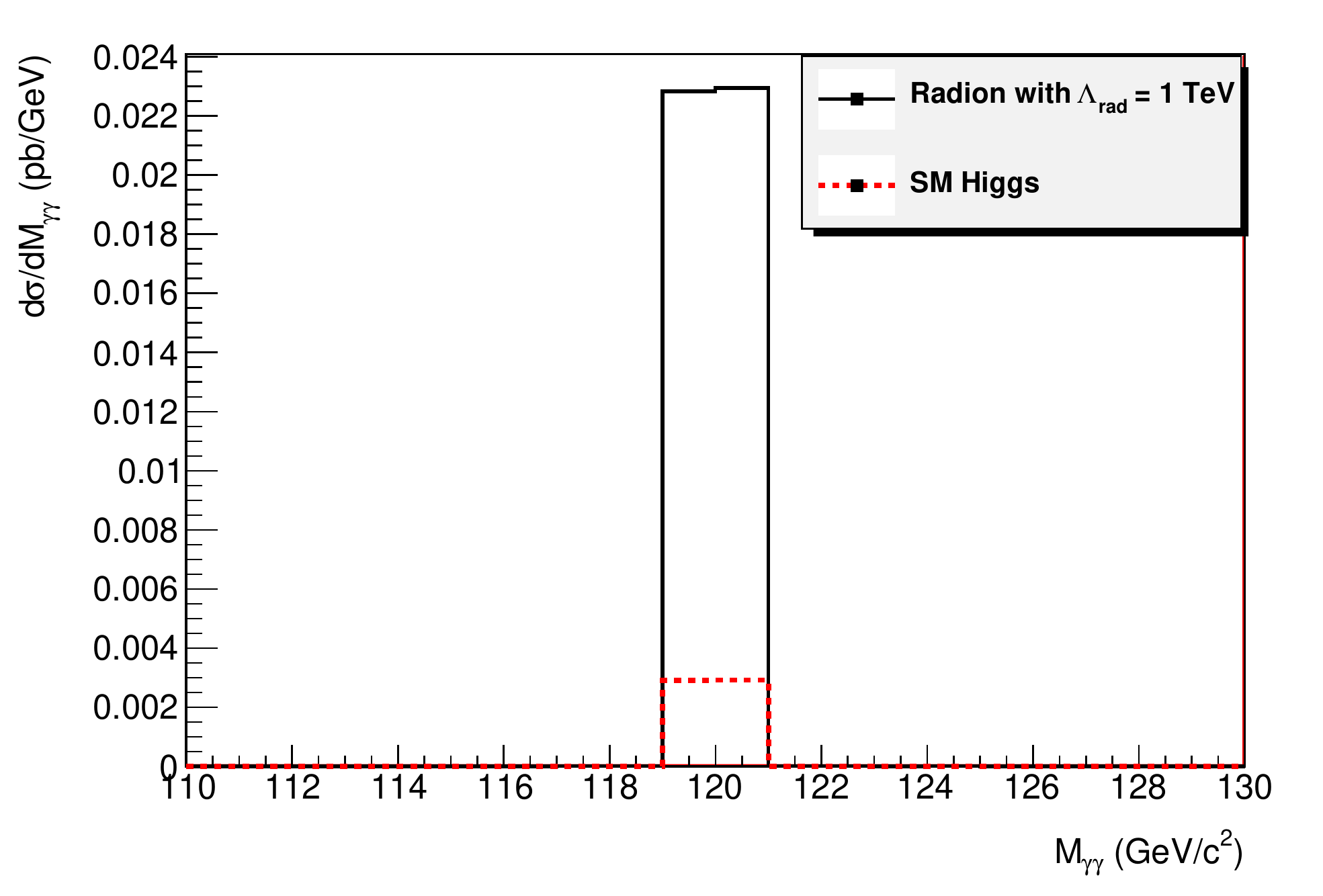}}
\caption{Left: contours of ratio of discovery significance for a radion vs. SM Higgs of the same mass, $R^{\gamma\gamma}_S$, in the plane of $m_{\rm rad}$-$\Lambda_{\rm rad}$. Right: $d\sigma/dM_{\gamma\gamma}$ for a radion vs. SM-Higgs of the same mass for $m_{\rm rad}=120\,{\rm GeV}$ and $\Lambda_{\rm rad}=1\,{\rm TeV}$.} 
\label{RadHcontourGamma}
\end{figure}

In order to compare the di-photon channel of the radion and the SM Higgs at the colliders,
it is useful to define an approximate formula for the ratio of the discovery significance of the
radion in the $gg\rightarrow r\rightarrow \gamma\gamma$ channel and that
of the SM-Higgs with the same mass, similar to the definition given in~\cite{GRW}, as
\begin{align}  \label{Eq:Rgg}
	R_S^{\gamma\gamma} \equiv\frac{S(r)}{S(h_{\rm SM})} 
=	\frac{\Gamma(r\rightarrow gg)\,B(r\rightarrow \gamma\gamma)}{\Gamma(h_{\rm SM}\rightarrow gg)\,B(h_{\rm SM}\rightarrow\gamma\gamma)}\ .
\end{align}
A similar ratio can be defined for $gg\to r \to \tau^+\tau^-$ and $gg\to r \to WW^*$ channels. 
Note that since both the SM Higgs and the radion are of narrow widths in the range of our interest, we do not include invariant mass resolution effects. 
By a careful examination of $R_S^{\gamma\gamma}$ we can see that the ratio of the widths weakly depends on the radion and the SM Higgs masses and it has a simple $\Lambda_{\rm rad}$ dependence  
\begin{align}
	\frac{\Gamma(r\rightarrow gg)}{\Gamma(h_{\rm SM}\rightarrow gg)} = \frac{v^2}{\Lambda^2_{\rm rad}}f_1(m_{\rm rad}) \ , \qquad
	\frac{B(r\rightarrow \gamma\gamma)}{B(h_{\rm SM}\rightarrow\gamma\gamma)} = f_2(m_{\rm rad}) \ ,
\end{align}
where $v\simeq246\,{\rm GeV}$ is the SM Higgs VEV and $f_1(m_{\rm rad})$ is monotonically changed by $\mathcal{O}(10\%)$ in the interval of $\mathcal{O}(100\,{\rm GeV})$ which covers the mass range we are interested in. 
$ f_2(m_{\rm rad})$ is a non-trivial function of $m_{\rm rad}$ and its value varies within $\mathcal{O}(1)$.
In Fig.~\ref{RadHcontourGamma} (LHS) we show that the signal of the radion can be easily enhanced by $\mathcal{O}(10)$ compared to the SM Higgs with the same mass. We assume that there is no Higgs-curvature coupling, and thus Higgs-radion mixing is absent. We also do not include brane-localized kinetic terms. 
In Fig.~\ref{RadHcontourGamma} (RHS) we show the differential cross-section, $d\sigma/dM_{\gamma\gamma}$, for the LHC (at center of mass energy of $7\,{\rm TeV}$) as a function of the di-photon invariant mass for a radion vs. SM Higgs of a mass of $m_{{\rm rad},h_{\rm SM}}=120\,{\rm GeV}$ and $\Lambda_{\rm rad}=1\,{\rm TeV}$ . The cross section is simulated by MadGraph/MadEvent  5.1 \cite{Alwall:2011uj,MGME} and we use CTEQ6M PDF set.

\begin{figure}[htbp]
\centerline{
	\includegraphics[width=.5\hsize]{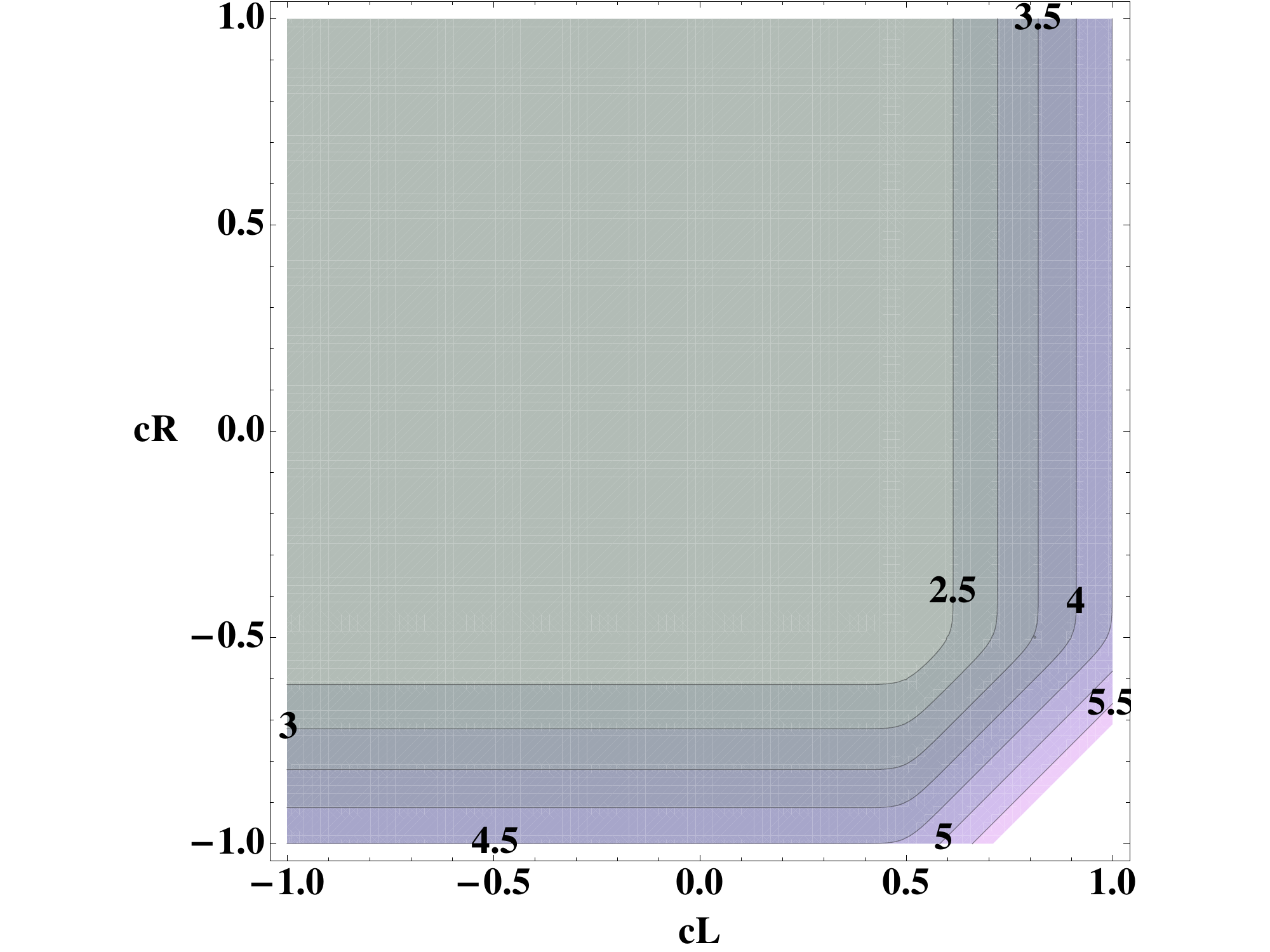}
	\includegraphics[width=.5\hsize]{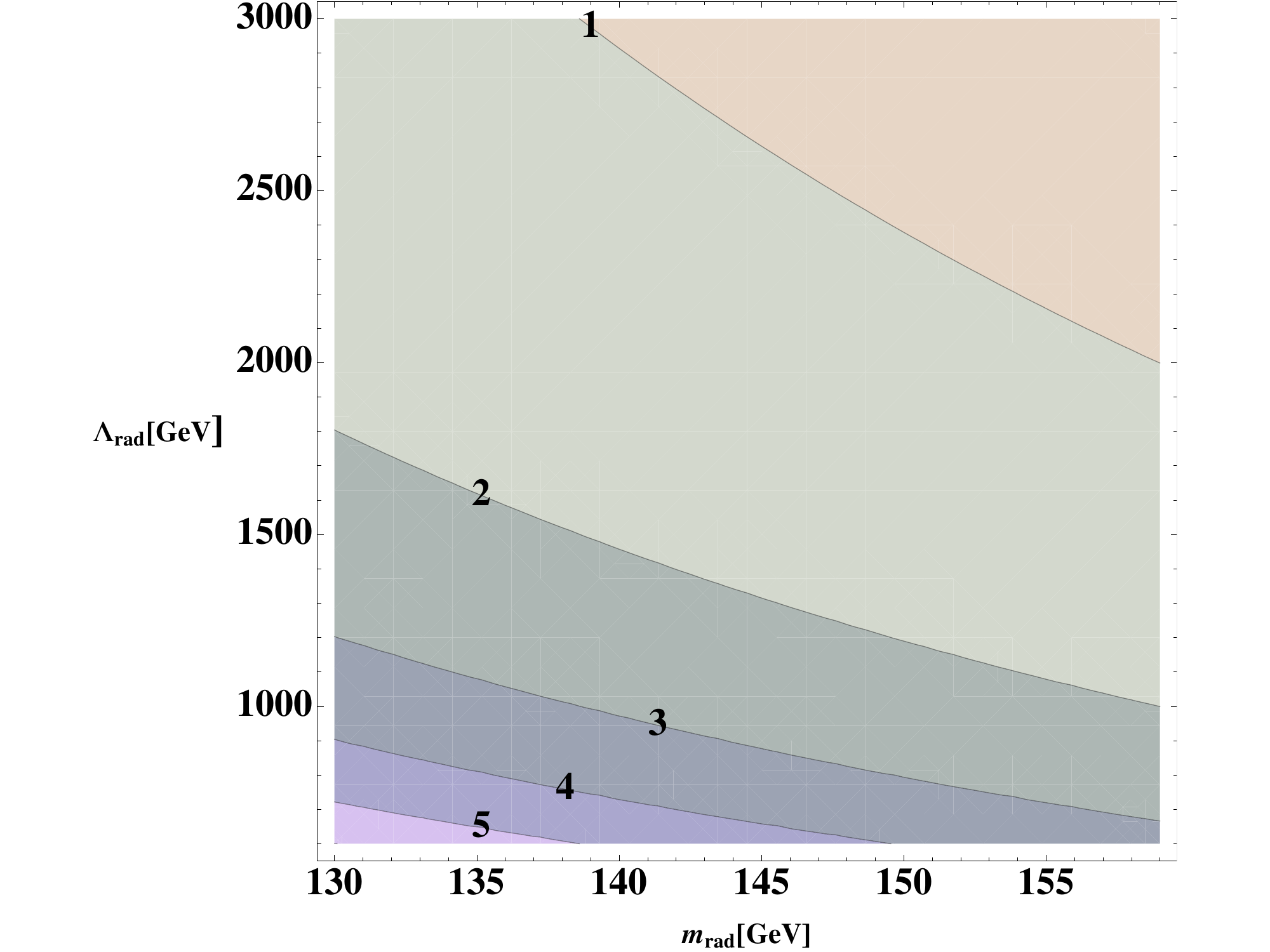}}
\caption{Left:  contours of ratio of discovery significance for a radion vs. SM Higgs of the same mass, $R^{\tau^+\tau^-}_S$ in the $c_L$-$c_R$ plane  for $m_{\rm rad}=120\,{\rm GeV}$ and $\Lambda_{\rm rad}=1\,{\rm TeV}$. Right: contours of ratio of discovery significance for a radion vs. SM Higgs of the same mass, $R^{WW^*}_S$, in the plane of $m_{\rm rad}$-$\Lambda_{\rm rad}$}
\label{RadHcontour}
\end{figure}

Another two interesting channels for the LHC are  $gg\to r\to\tau^+\tau^-$ and $gg\to r\to WW^*$, where the discovery significance is defined similar to the one of $\gamma\gamma$ in Eq.~\eqref{Eq:Rgg} and are labeled $R^{\tau^+\tau^-}_S$ and $R^{WW^*}_S$ respectively. 
The general formula for the coupling of the radion to the SM-fermions with the bulk mass parameters $c_L$ and $c_R$, defined in \cite{CHS}, is
 \begin{align}
	\frac{1}{2}\frac{m_f}{ \Lambda_{\rm rad}}\frac{  (1+2 c_L)+ a_\star^{2 c_L+2 c_R} (1-2 c_R)+2a_\star^{1+2 c_R} (c_R-c_L)-2a_\star^{2 c_L-1}}{\left(1-a_\star^{2c_L-1}\right) \left(1-a_\star^{1+2c_R}\right)}.
\end{align}
In Fig.~\ref{RadHcontour} (LHS) we show that the signal of the radion can be easily enhanced by $\mathcal{O}(5)$ compared to that of the SM-Higgs for a given mass. As we can see the signal significance is very sensitive to the localization of both LH and RH $\tau$'s in the bulk, i.e. $c_L$ and $c_R$.  In Fig.~\ref{RadHcontour} (RHS) we show that the signal of the radion to $WW^*$ can be bigger  by a factor of $\mathcal{O}(5)$ compared to the SM Higgs signal.

In a recent CDF study there is 4.1$\sigma$ excess in the $Wjj$ event sample in $7.3\,{\rm fb^{-1}}$ of data for dijet invariant masses between $120-160\,{\rm GeV}$, which corresponds to a sizable cross section of $4\,{\rm pb}$ \cite{CDFWjj,CDFWjjweb}. However, no excess was found in a similar study done by the D0 collaboration \cite{Abazov:2011af}.
Although the radion mass can be naturally in the range to explain this anomaly, we find that it is unlikely to account for it.
Below, we analyze the case where the SM fields propagate in the bulk. It is worth emphasizing that this analysis holds also for the original RS model (where the SM is IR-brane localized). This is due to the fact that the correction to the $W^+W^-$-radion coupling is about only 20\% compared to the bulk SM case \cite{CHS} and therefore will not change our conclusion.

The dominant channel for $Wjj$ production that involves the radion is Higgs-like radion-strahlung. Other channels are proportional to light quark masses and are negligible.  
Since the phenomenology of the radion and the SM Higgs are similar, it is easy to estimate the ratio between the rate for $p\bar{p}\to Wr \to Wjj$ (dominanted by $jj=gg$) and $p\bar{p}\to Wh_{\rm SM}\to Wjj$ (dominanted by $jj=b\bar b$), which it turns out to be 
\begin{align}
	\frac{\sigma(p\bar{p}\to Wr)BR(r\to jj)}{\sigma(p\bar{p}\to Wh_{\rm SM})BR(h_{\rm SM}\to jj)} \sim r_w\left(\frac{v}{\rm TeV}\,\frac{\rm TeV}{\Lambda_{\rm rad}}  \right)^2 \sim 0.1 \ ,
\end{align}
where  $r_w\simeq1.4$ comes from the sub-leading corrections to the $W^+W^-$-radion coupling, with $v/\Lambda_{\rm rad}\sim1/4$ and $BR(r\to jj)/BR(h_{\rm SM}\to jj)\sim 1$.
Explicit MadGraph/MadEvent simulation leads to  $\sigma{(p\bar{p}\to Wr \to Wjj)}\sim 1.1\, {\rm fb}$, while $\sigma(p\bar{p}\to Wh_{\rm SM})BR(h_{\rm SM}\to b\bar b)\sim 12\,{\rm fb}$ \cite{Djouadi:1997yw}, which is consistent with our estimation. 
Therefore, we conclude that the $Wjj$ anomaly can not be explained by the RS radion (contrary to the statement of \cite{Bhattacherjee:2011yh}).  

%%%%%%%%%%%
\section{Conclusions} \label {Conclusions}
%%%%%%%%%%%

Flavon stabilization of warped extra dimension form a unique mechanism, originally proposed by Rattazzi and Zaffaroni, where solutions to the gauge hierarchy problem and the new physics flavor problem are unified.
As we have shown in this setup, the stabilization is induced by a combination of flavon vacuum expectation values and the presence of IR-brane tension term. 
Parametric enhancement of the radion mass is induced by the brane vacuum energy.
It scales as $m_{\rm rad}^2\sim k^2a_\star^2\epsilon$, in agreement with \cite{KNQ},  where $\epsilon\sim1/\log(a_\star)\sim1/40$, $a_\star$ corresponds to the weak-Planck hierarchy and $ka_\star\sim\mathcal{O}({\rm TeV})$. 
Therefore, the natural range for the radion mass is $\mathcal{O}(100\,{\rm GeV})$ which is in the favored range of the standard model (SM) Higgs mass.
Note that the radion mass is large enough to avoid radion-mediated flavor changing neutral currents~\cite{ATZ}, but still small enough to induce a sizable Sommerfeld enhancement that contributes to the dark matter (DM) annihilation, which may be probed via indirect astrophysical signals for TeV scale dark matter in Randall-Sundrum models~\cite{ABSP}~\footnote{In RS framework, the first DM model was based on a
grand unified theory (GUT) model~\cite{Agashe:2004ci,Agashe:2004bm}, where stability of the DM is a spin-off of suppressing proton decay, and later extended to incorporate custodial $Z\to b\bar{b}$ into the GUT frame work~\cite{ABSP}. It is clear that radion mediated Sommerfeld enhancement might also be relevant for {\em other} RS-type scenarios with a DM candidate~\cite{Agashe:2007jb,Panico:2008bx,Cui:2009xq}, with a exception for~\cite{Medina:2011qc}, where DM candidate is KK-odd radion.}.

In this mass range the di-photon channel, $gg\rightarrow r/h_{\rm SM}\rightarrow \gamma\gamma$, forms an important discovery channel for both the radion and the SM Higgs, which tend to have similar collider phenomenology. We have emphasized that in this channel the radion signal can be naturally enhanced by $\mathcal{O}(10)$ compared with that of the Higgs. 
We also pointed out that in two other interesting channels, $gg\rightarrow r/h_{\rm SM}\rightarrow \tau^+\tau^-$ and $gg\rightarrow r/h_{\rm SM}\rightarrow WW^*$, the radion signal is similarly enhanced by a factor of $\mathcal{O}(5)$.
We find that the radion is unlikely to account for anomalies in the $W$ plus dijet differential distribution.

%%%%%%%%%%%%%%
\section*{Acknowledgments}
%%%%%%%%%%%%%%
We thank Kaustubh Agashe, Ofer Aharony, Guy Gur-Ari, Nizan Klinghoffer, Zohar Komargodski, Thomas Konstandin, Sreerup Raychaudhuri, Adam Schwimmer and Geraldine Servant for useful discussions. GP is the Shlomo and Michla Tomarin career development chair and is supported by the
Israel Science Foundation (grant \#1087/09), EU-FP7 Marie Curie, IRG fellowship, Minerva and G.I.F., the German-Israeli
Foundations, and the Peter \& Patricia Gruber Award.

% All the sections from here will be Appendix of the paper
%%%%%%%%%
\appendix %%%
%%%%%%%%%

%%%%%%%%%%%%%%%%%%%%%%%%%
\section{Derivation of the Radion Effective action} \label{Deriv_R4DEA}
%%%%%%%%%%%%%%%%%%%%%%%%%

Here we derive the radion effective action following the method of \cite{GW1,GW2}. The radion is identified with a simple field ($x^\mu$ dependent) parameterization of the fifth dimension radius.
The fifth dimension coordinate is transformed into polar coordinate, $y=T(x^\mu)\theta=T\theta$ and $\theta\in(-\pi,\pi]$, where the line element is given by
\begin{align}
&	ds^2=\eexp{-2A(T\abs{\theta})}\eta_{\mu\nu}dx^\mu dx^\nu - T^2 d\theta^2 \ . \label{Tfluc} 
\end{align}
and $A(T\abs{\theta})$ is given by Eq.~\eqref{Asol}. The 4D-effective action is derived by integrating out the fifth dimension
\begin{align} \label{L_4D_GW_int}
	\mathcal{L}^{\rm 4D}_{\rm eff,GW} 
=	 \int_{-\pi}^\pi d\theta \left\{\mathcal{L}_{\rm gravity}\left[T\theta\right] 
		+ \sum_i \mathcal{L}^{(i)}_{\rm GW}\left[\phi_i\left(T\theta\right)\right]  \right\} \ .
\end{align}
where
\begin{align}
&	\mathcal{L}^{(i)}_{\rm GW} 
= 	 \frac{T\eexp{-4A}}{2}\left\{\eexp{2A}(\partial_\mu\phi_i)^2-\phi'^2_i- \epsilon_i k^2\phi_i^2 \nonumber 
	-\frac{k}{T}\left[ b_{\rm IR,i}\left(\phi_i - r\nu_{{\rm IR},i} M^{3/2}\right)^2\delta(\theta-\pi) \right]  \right\}  , \nonumber\\
&	\mathcal{L}_{\rm gravity}
= 	M^3T\eexp{-4A}\left\{12k^2-\mathcal{R} -\frac{k}{T}\left[ \delta(\theta)T_{\rm UV}+\delta(\theta-\pi)T_{\rm IR}\right] \right\} \ , \nonumber\\
&	\mathcal{R}T\eexp{-4A} 
=	6\eexp{-2A}\left( \theta^2TA'^2-\abs{\theta}A' \right)(\partial_\mu T)^2
	+4\eexp{-4A}\left[5TA'^2-2TA''-4A'\left(\delta(\theta)-\delta(\theta-\pi)\right) \right] \ . \nonumber
\end{align}
To order $r^2$ and leading order in $a_\star$ and $\epsilon$ (note that we do not expand terms such as $\eexp{-kT\pi\epsilon}$ because $k\pi\epsilon\langle T \rangle\sim\mathcal{O}(1)$),  by using Eqs.~\eqref{Azero_sol} and~\eqref{af_def} the 4D effective action is given by
\begin{align} 
	\mathcal{L}^{\rm 4D}_{\rm eff,GW} 
=&	\frac{1}{2} f^2 \left(1+\frac{r^2}{12}\tilde{h}\left(a\right) \right)\left(\partial_\mu a\right)^2-V_{\rm eff}(a) +const \ .
\end{align}
where $V_{\rm eff}(a)$ is given by Eq.~\eqref{VGW} and
\begin{align} \nonumber
	\tilde{h}\left(\eexp{-k\pi T}\right) 
=	 \int_{0}^\pi\frac{d\theta}{\pi^2} \eexp{2kT(\pi-\theta)} \left\{ \frac{T}{kM^3}\left(\frac{1}{r}\pd{\phi_i}{T}\right)^2
		+\theta\left[2k(kT\theta-1)G+(1-2 k T \theta )G'\right]\right\} \ .
\end{align}

Since we look for a solution around the VEV of $a$, denote $\langle a \rangle = a_\star\ll1$ which is $x^\mu$ independent, $\tilde{h}(a)$ can not change the value of $a_\star$. Therefore, we plug $a_\star$ into $\tilde{h}(a)$\,\footnote{For small $\epsilon$, the numerical value of $\tilde{h}(a_\star)$ is of order unity.} and omit all the interaction terms which are proportional to $\partial_\mu a$. Schematically, the 4D action is of the form
\begin{align}
	\mathcal{L}^{\rm 4D}_{\rm eff,GW} 
=	\frac{1}{2}f^2\left(1+\frac{r^2}{12}\tilde{h}(a_\star) \right)\left(\partial_\mu a\right)^2 - r^2\bar{V}(a) \ ,
\end{align}
where $r^2\bar{V}(a)=V_{\rm eff}(a)$. Moving to a canonical basis where the appropriately normalized radion field is denoted by $\varphi_R$, $\mathcal{L}^{\rm 4D}_{\rm eff,GW}$ is given by
\begin{align}
	\mathcal{L}^{\rm 4D}_{\rm eff,GW}
=&	\frac{1}{2}\left(\partial\varphi_{\rm R}\right)^2 - r^2\bar{V}\left(\frac{\varphi_{\rm R}}{f\sqrt{1+\frac{r^2}{12}\tilde{h}(a_\star)}}\right)  \nonumber\\
=&	\frac{1}{2}\left(\partial\varphi_{\rm R}\right)^2 -  r^2 \bar{V}\left(\frac{\varphi_{\rm R}}{f} \right) + \mathcal{O}(r^4) \ .
\end{align}
As one can see the contribution of $\tilde{h}(a_\star)$ vanishes at order $r^2$. The canonical form of the radion field is
\begin{align}
	\varphi_{\rm R}(x^\mu) = fa=\sqrt{\frac{12M^3}{k}}\eexp{-k\pi T(x^\mu)}  \ .
\end{align}

%%%%%%%%%%%%%%%%%%%%%%
\section{Goldberger-Wise and Casimir Energy Stabilizations} \label{Casimir_potential}
%%%%%%%%%%%%%%%%%%%%%

A different approach from GW-mechasim to the radion stabilization is to implement the Casimir energy which is induced by bulk fields \cite{GP,GR,GPT}. 
Although the Casimir energy looks as a natural candidate for stabilization because one does not need to introduce a special field for stabilization (as in GW mechanism), it is unnatural since it involves a small parameter, $\mathcal{O}(10^{-5})$ or smaller, in order to get the right hierarchy \cite{GR,GPT}. This small parameter induces a fine-tuning, and therefore this mechanism can not be viewed as a complete solution to the hierarchy problem.  
However, in models that contains GW mechanism and bulk fields, one can ask if the Casimir energy ruins or reinforces the stabilization caused by the GW mechanism.

In order to address this question we consider the GW-stabilization and the dominant contributions to the Casimir energy\footnote{Bulk scalars with mass $m^2_{s}=-4k^2$ and bulk fermions with bulk mass, $m_{f}=k/2$, induce similar contribution to the Casimir energy \cite{GP,GR}. Since such scalars do not appear in RS phenomenologically viable models and the fermion bulk masses are rather model dependent, we neglect their contribution to the Casimir energy. However, bulk gauge fields do appear in models were the SM-fields are allowed to propagate in the bulk, therefore we consider their contribution to the Casimir energy.}
 from bulk gauge fields. The Casimir energy due the bulk gauge fields were derived in \cite{GP}. In the case of one GW scalar and $N$ bulk gauge fields the radion 4D-effective potential is:
\begin{align} 
	V_{\rm eff}(a)
=	r^2kM^3a^4  \Bigg\{&\delta{T}_{\rm IR} + \frac{4b_{\rm IR}}{8+b_{\rm IR}} (a^{\epsilon/4}-\nu )^2
		+\frac{\epsilon}{4(8+b_{\rm IR})^2}\Big[b_{\rm IR}^2\nu ^2 \nonumber\\
&		+2a^{\epsilon /2}\left(b_{\rm IR}^2-32\right) - 4a^{\epsilon /4}\nu b_{\rm IR}(4+b_{IR})  \Big] \Bigg\}
	+\frac{k^4N_g\beta(\rho_{\rm IR}) a^4}{16\pi^2\log(a)}\ , \label{V_Casimir}
\end{align}
where
\begin{align}
&	\beta(\rho_{\rm IR}) \equiv -\int_0^\infty dt t^3\left( \frac{t^2\rho_{\rm IR}K_1(t)-tK_{0}(t)}{t^2\rho_{\rm IR}I_1(t)+tI_{0}(t)} \right)  \ . \nonumber
\end{align}
$K_i(t)$ and $I_i(t)$ are Bessel functions, $N_g=g_pN$, $g_p$ is the number of the physical polarizations ($g_p=3$ for 5D bulk gauge fields) and $\rho_{\rm IR}$ is the coefficient of the IR-brane kinetic term of the gauge fields. $\beta(\rho_{\rm IR})$ is an $\mathcal{O}(1)$ parameter and its sign depends on the value of $\rho_{\rm IR}$ (for example $\beta(0)=1.005$, $\beta(5)=-1.519$ and $\beta(\infty)=-2.330$).  
The extremum and the radion induced mass from the potential in Eq.~\eqref{V_Casimir}, to leading order in  $\epsilon$ and $1/\log(a_\star)$, are identical to the ones without the Casimir energy and given in Eqs.~\eqref{a_term}-\eqref{mrad}.
 Therefore, we conclude that the Casimir energy has only sub-leading effect on the radius stabilization. 

In \cite{GPT} it is mentioned that the Casimir energy does not affect the radion stabilization due to GW mechanism, which is in agreement with our findings. The difference between the two works, is that in \cite{GPT} the Casimir energy is due to the presence of a conformally massless bulk scalar (since conformal invariance is assumed). The effective potential in \cite{GPT} behaves as $V(a)\sim k^4a^4\left(1 - da \right) +\mathcal{O}(a^6)$. Stabilization with this potential leads to a tiny mass of the radion $m^2_{\rm rad}\sim k^2a^3_\star$ which has no effect on the GW-stabilization. As with the other stabilization mechanisms based on the Casimir effect, it suffer from a fine-tuning problem, \cite{GPT}.

%%%%%%%%%%%%%%%%%%%%%%%%%%%%%
\section{The Radion Mass from Linearized Einstein Equations} \label{linearied_EE}
%%%%%%%%%%%%%%%%%%%%%%%%%%%%%%

In this appendix we derive the radion mass in the method of \cite{CGK}, which is extended by \cite{KMP}. Here we describe trivial extension of it to the flavon case, where there is more than one bulk scalar.  
Unlike the naive ansatz described in Appendix~\ref{Deriv_R4DEA}\,, this method takes into account not only the metric's fluctuations but also the bulk scalars' fluctuations. In addition, both scalars and metric fluctuations are solutions to the linearized Einstein equations.
The parametrization used for the radion and the bulk scalars fluctuations is   
\begin{subequations}
\begin{align}
	&ds^2=e^{-2A(y)-2 F(x,y)}\eta_{\mu\nu}dx^\mu dx^\nu  -(1+G(x,y))^2dy^2 \,,\\
	&\tilde{\phi}_i(x,y)=\phi_i(y)+\chi_i(x,y)\,,
\end{align}
\end{subequations}
where $A(y)$ and $\phi_i(y)$ are the background solution, see Eqs.~\eqref {GW_background} and \eqref{Asol}.
One can add graviton fluctuations by replacing $\eta_{\mu \nu} \to \eta_{\mu \nu}+ h_{\mu \nu}^{TT}$, were $TT$ denotes transverse traceless. However, as one can check, these graviton fluctuations will be decoupled in the linearized Einstein equations (see also \cite{AG}). 
Note that in this section we use the notation of \cite{CGK}, denoting the bulk potential as $V$ and the brane potentials as $\lambda_{\rm UV,IR}$. The bulk cosmological constant and brane tensions are included in the bulk potential and brane potential respectively. 
From the $\mu\neq \nu$ components of the Einstein equations one finds the equation
\begin{align}
	 2\partial_\mu\partial_\nu F - \partial_\mu\partial_\nu G=0 \,.
\end{align}
Therefore, we set $G=2F$ from here onwards. This scalar degree of freedom, $F$, is the radion\,\footnote{\, In fact we identify only lowest mass eigenstate as the radion: without the GW mechanism there is only one solution with zero mass; with the GW there is a KK tower of solutions for the metric-scalar fluctuations. For details see \cite{CGK}.} and its mass is identified with the lowest eigenvalue of the 4D box operator, $\square=\eta^{\mu\nu}\partial_\mu\partial_\nu$, 
\begin{align} \label{BOX}
	\square F = -m^2 F \,.
\end{align} 

The linearized Einstein equations, $\delta R_{M N} = \kappa^2 \delta \tilde{T}_{M N}$, where $\kappa^2=1/(2M^3)$.
$\delta R_{M N}$ is the linearized Ricci tensor and $\delta \tilde{T}_{M N}$ denotes linearization of the tensor $\tilde{T}_{MN}=T_{M N}-\frac{1}{3} g_{MN} g^{ CD} T_{CD}$ built from the energy-momentum tensor $T_{MN}$. Their components are
\begin{subequations}
\begin{align}
	&\delta R_{\mu \nu} = \eta_{\mu \nu} \square F + e^{-2A} \eta_{\mu \nu} (-F'' + 10 A' F' + 6 A'' F -24 A'^2 F ) \,,  \\
	&\delta R_{\mu 5} = 3 \partial_{\mu} F' - 6 A' \partial_{\mu} F \,,  \\
	&\delta R_{55} = 2 e^{2 A} \square F + 4 F'' - 16 A' F' \,,  
\end{align}
\end{subequations}
and
\begin{subequations}
\begin{align}
	&\delta \tilde{T}_{\mu \nu} = - \frac{e^{-2A}}{3}\!\left[ 2V_i(\{\phi\}) \chi_i - 4 V(\{\phi \}) F 
		+ \sum_{\alpha}\left( {\lambda_{\alpha}}_i(\{ \phi \}) \chi_i -4 \lambda_{\alpha}(\{\phi \})F \right)\delta(y-y_{\alpha})  \right]\!\!\eta_{\mu \nu}  , \\
	& \delta \tilde{T}_{\mu 5} = \phi'_i \partial_{\mu} \chi_i  \,, \\
	& \delta \tilde{T}_{55} = 2 \phi'_i \chi_i'+ \frac{2}{3} V_i (\{ \phi\}) \chi_i+ \frac{8}{3} V(\{\phi\}) F 
		+ \frac{4}{3} \sum_{\alpha} \left({\lambda_{\alpha}}_i (\{\phi \})\chi_i + 2 \lambda_{\alpha} (\{\phi \}) F \right) \delta y- y_{\alpha}) \,, 
\end{align}
\end{subequations}
where we use the notation $V_i (\{\phi\})\equiv \frac{\partial V}{ \partial \tilde{\phi}_i}\big|_{\tilde{\phi}_j=\phi_j}$,  $V_{ij} (\{\phi\})\equiv \frac{\partial^2 V}{ \partial \tilde{\phi}_i \partial \tilde{\phi}_j}\big|_{\tilde{\phi}_k=\phi_k}$ and similarly for $\lambda_{\alpha}$; repeated indices are summed over unless otherwise mentioned. The linearized scalar equations of motion are
\begin{align}
	e^{2 A} \square \chi_i - \chi_i'' + 4A' \chi_i' + V_{ij}(\{ \phi \}) \chi_j=&-6 \phi_i' F' - 4 V_i (\{\phi \}) F  \nonumber \\
	&- \sum_{\alpha} \left[{\lambda_{\alpha}}_{ij}(\{\phi \}) \chi_{j} +2 {\lambda_{\alpha}}_{i}(\{\phi \}) F \right]\delta(y-y_{\alpha})\,.
\end{align}
The $\mu 5$ equation can be immediately integrated, setting the integration constant to zero 
\begin{align} \label{5muint}
	3 F' - 6 A' F = \kappa^2 \phi_i ' \chi_i \,.
\end{align}

The boundary conditions for the scalars and radion fluctuation are obtained by integrating infinitesimally around the delta functions of the $\mu \nu$ equation and $\chi_i$ equation of motion. After use of the background jump conditions\,\footnote{\, In the notation used here, these are $[A']\big|_{\alpha}= \frac{\kappa^2}{3} \lambda_{\alpha}(\{\phi\})$ and $[\phi_i ']\big|_{\alpha}= {\lambda_{\alpha}}_i(\{\phi\})$.} these are
\begin{align}
	 &[F'] \big|_{\alpha}=   \frac{\kappa^2}{3}  \left( {\lambda_{\alpha}}_i(\{ \phi \}) \chi_i +2 \lambda_{\alpha}(\{\phi \})F \right) \,, \\
	 &[ \chi_i' ]\big|_{\alpha}= {\lambda_{\alpha}}_{ij}(\{\phi \}) \chi_{j} +2 {\lambda_{\alpha}}_{i}(\{\phi \}) F  \,. \label{chibc}
\end{align}
Using the background jump conditions the first equation is equivalent to Eq.~\eqref{5muint} and therefore does not constrain the solution. 

As in \cite{CGK} an eigenvalue equation for $F$ is now formed. This is first done by considering the combination of $e^{2A} (\mu ,\nu) + (5, 5)$ in the bulk, where $(A, B)$ denotes the $AB$ component of the linearized Einstein equation. After use of Eq.~\eqref{BOX} one gets
\begin{align}
	  F'' -2 A' F' + 2 A'' F -8 A'^2 F -  e^{2A} m^2 F =\frac{\kappa^2}{3}( 2 \phi'_i \chi_i'+ 4 V(\{\phi\}) F ) \,.
\end{align}
Using Eqs.~\eqref{EEq} in the bulk further simplifies this to
\begin{align}
	 F'' -2 A' F' - e^{2A} m^2 F =\frac{2\kappa^2}{3} \phi'_i \chi_i' \,.
\end{align}
In the case of a single scalar field one can now use Eq.~\eqref{5muint} to get the eigenvalue equation for F
\begin{align}
		F'' - 2A'F' -4A''F-2\frac{\phi''}{\phi'}F' + 4A'\frac{\phi''}{\phi'}F = -m^2e^{2A}F  \,.\label{Feq}
\end{align}
In the case of many scalar fields this elimination is not easily achieved and one generally has to diagonlize the entire system (see e.g. \cite{AG} for such attempts). However, for the case considered in section \ref{shinstab} we can use the fact that the $\phi_i$'s have scaled profiles, i.e. $\phi_i=r \nu_{{\rm UV},i} p(y)$, where $p(y)$ is general for all the flavons (with $\nu_{{\rm UV},i}=0$ for the fields with vanishing background profiles). In that case we can rewrite Eq.~\eqref{5muint} as
\begin{align} \label{5muchip}
	\frac{3}{\kappa^2 }\left( F' - 2 A' F \right) = r \nu_{{\rm UV},i} p'(y) \chi_i \,.
\end{align}
Therefore,
\begin{align} 
	 \phi_i ' \chi_i ' =r \nu_{{\rm UV},i} p' \chi_i '=\frac{3}{\kappa^2} \left( - \frac{p''}{p'}(F'-2A'F) +F'' -2 A'' F- 2A' F' \right) \,,
\end{align}
and using this, one gets Eq.~\eqref{Feq} with $\phi$ replaced by $p$. As in \cite{CGK}, to find the lightest mode, one solves Eq.~\eqref{Feq} perturbatively in $r^2$, using the ansatz 
\begin{align}
	F(x,y)=e^{2ky}\left(1+r^2f(y)\right)\tilde{\varphi}_r(x) \,, && m^2=0+ r^2 \tilde{m}^2 \,.
\end{align}
Note that this ansatz is justified by noting that if we neglect $A''$ in the bulk then $F=e^{2 A}$ is a solution of zero mass. Denoting 
\begin{align}
	Q(y)=\frac{\phi''}{\phi'}=\frac{p''}{p'}\,, && A(y)=k y + r^2 k G(y)   \,,
\end{align}
 and expanding the Eq.~\eqref{Feq} to order $r^2$, one gets
\begin{align} \label{smallfeq}
		f''(y)  + 2(k-Q(y))f'(y) -4k(k-Q(y))G'(y)  - 4kG''(y) + e^{2ky} \tilde{m}^2= 0 \,,
\end{align}
where the boundary conditions  given by Eq.~\eqref{chibc}.
In the case of infinite brane masses the boundary conditions are simpler and by use of Eq.~\eqref{5muint} imply
\begin{align} \label{infbc}
	\left(F'-2 A' F \right) \big|_{\alpha}=0\,.
\end{align}  
Expanding this boundary conditions to order $r^2$ yields 
\begin{align}
	f'(y_\alpha) = 2kG'(y_\alpha)\ .
\end{align}
The bulk equation Eq.~\eqref{smallfeq} has two integration constants and the mass parameter, one integration constant is an overall normalization factor, while the mass and the other integration constant are found by imposing the boundary conditions. 
We have solved Eq.~\eqref{smallfeq} and found the radion mass to lowest order in $\epsilon$, $r$ and $a_{\star}$ agree \emph{exactly} with the masses calculated in the effective potential in sections~\ref{GWStabMech} and~\ref{shinstab}\,, i.e. Eqs.~\eqref{mrad} and~\eqref{shinmrad}.

%%%%%%%%
% references %
%%%%%%%%
\bibliographystyle{utcaps}
\bibliography{references}

\end{document}